\documentclass[aps,pre,twocolumn]{revtex4-1}
\usepackage[bookmarks,colorlinks,citecolor=blue,linkcolor=blue]{hyperref}
\usepackage{amssymb}
\usepackage{amsmath,bm}
\usepackage{graphicx,color}
\usepackage{bbm}
\usepackage[bottom]{footmisc}
\usepackage{multirow}
\newcommand{\B}[1]{{\bm{#1}}}

\begin{document}
\title{Unifying homogeneous and inhomogeneous rheology of dense suspensions}
\author{Bhanu Prasad Bhowmik}
\email{bhowmikbhanuprasad592@gmail.com}
\author{Christopher Ness}
\email{chris.ness@ed.ac.uk}
\affiliation{School of Engineering, University of Edinburgh, Edinburgh EH9 3JL, United Kingdom}

\begin{abstract}
{The rheology of dense suspensions lacks a universal description due to the involvement of a wide variety of parameters, ranging from the physical properties of solid particles to the nature of the external deformation or applied stress. While the former controls microscopic interactions, spatial variations in the latter induce heterogeneity in the flow, making it difficult to find suitable constitutive laws to describe the rheology in a unified way. For homogeneous driving with a spatially uniform strain rate, the rheology of non-Brownian dense suspensions is well described by the conventional $\mu(J)$ rheology. However, this rheology fails in the inhomogeneous case due to non-local effects, where the flow in one region is influenced by the flow in another. Here, motivated by observations from simulation data,  we introduce a new dimensionless number, the suspension temperature $\Theta_s$, which contains information on local particle velocity fluctuations. We find that $\mu(J,\Theta_s)$ provides a unified description for both homogeneous and inhomogeneous flows. By employing scaling theory, we identify a set of constitutive laws for dense suspensions of  frictional spherical particles and frictionless rod-shaped particles. Combining these scaling relations with the momentum balance equation for our model system, we predict the spatial variation of the relevant dimensionless numbers, the volume fraction $\phi$, the viscous number $J$, the macroscopic friction coefficient $\mu$, and $\Theta_s$ solely from the nature of the imposed external driving.}
\end{abstract}

\maketitle

\section{Introduction}
Dense suspensions consist of Brownian or non-Brownian solid particles suspended in viscous fluids in roughly equal proportions. Such materials, for example, corn starch in water, slurries, blood, etc., have widespread applications in both daily life and industry, and their flow is observed in many natural phenomena. Therefore, understanding their rheology is pivotal~\cite{NessAnnualReview2022,StickelAnnualReview,Guazzelli_Pouliquen_2018,Houssais2015}. However, the rheology of these materials is extremely complex and lacks a universal description due to its dependence on various parameters such as the size, asphericity, and surface smoothness of the solid particles, the nature of the suspending fluid, as well as the complexity involved in the external driving. Various constitutive models have been introduced based on defining appropriate dimensionless numbers, but their validity is often limited to specific scenarios~\cite{MansardAndColinSoft2012,BoyerGuazelliPouliquenPRL2011,Tapia2022,ConstLaw1,ConstLaw2,ConstLaw3}. Therefore, identifying suitable quantities to establish constitutive laws to describe the rheology of different types of dense suspensions under various driving conditions has been a topic of intense research~\cite{GillissenNessPRL2020, BoyerGuazelliPouliquenPRL2011, TigheNonLocal2019,BPBNessPRL2024,KimKamrin2020PRL,PahtzPRL2019,LernerWyartPRE2015,GuazzelliPRF2024}.

Under the homogeneous scenario, where the strain rate is spatially uniform, the rheology of the dense suspensions is well described by three dimensionless quantities: the solid volume fraction $\phi$; the ratio of the viscous time scale $\eta/P$ to shear time scale $1/\dot{\gamma}$ known as viscous number $J=\eta\dot{\gamma}/P$; and the ratio of shear stress $\sigma_{xy}$ to normal stress $P$ known as effective or macroscopic friction coefficient $\mu = \sigma_{xy}/P$ ~\cite{BoyerGuazelliPouliquenPRL2011}. Here $\eta$ and $\dot{\gamma}$ are suspending fluid viscosity and strain rate, respectively. The interdependence of these dimensionless numbers forms the constitutive laws for homogeneous rheology of dense suspensions. However, the validity or structure of such constitutive laws might alter depending on the various properties of the constituent particles. Nevertheless, in real systems the nature of the rheology is often inhomogeneous due to the spatial variation of the strain rate, so the aforementioned dimensionless numbers are not sufficient~\citep{GillissenNessPRL2020}. Under homogeneous straining, $J$ decreases with decreasing $\mu$ and increasing $\phi$, and eventually vanishes when $\mu$ and $\phi$ approach their limiting value $\mu_J$ and homogeneous shear jamming volume fraction $\phi_J^H$, respectively. This physically signifies the cessation of the flow, however, in the case of inhomogeneous flow one can observe $\phi>\phi_J^H$ and $\mu <  \mu_J$ even for finite $J$ due to shear-induced particle migration~\cite{GADALA-MARIA1979,KARNIS1966531,migration1, migration2,migration3,Fall2010,MATASMORRISGUAZZELLI2004} and the non-local effect~\cite{PouliquenForterre2009,TigheNonLocal2019,KimKamrinNonLocal2023,Bouzid2015,NottBrady1994}.

Non-local phenomena in the rheology of various soft matter is studied extensively, and pictured as a process where flow in the regions with $\mu > \mu_J$ facilitates the flow in regions with $\mu < \mu_J$ via diffusion of local fluidity of the system~\cite{Goyon2008}. Such diffusion of local fluidity can be described by an inhomogeneous Helmholtz-like equation which suggests a cooperative motion controlled by an inherent length scale ~\cite{BocquetPRL2009,BouzidTrulssonPRL2013}. In the context of dry granular systems, in Ref.~\cite{KamrinKovalPRL2012} granular fluidity is macroscopically defined as $g = \dot{\gamma}/\mu$. Later in Ref.~\cite{ZhangKamrinPRL2017}, Zhang and Kamrin established the microscopic definition of fluidity, denoted as $\tilde{g}$, in terms of fluctuations of particle velocity, $\delta u$. $\tilde{g}$ is uniquely determined by the local $\phi$, and can be expressed as $\tilde{g} = g a/\delta u = F\left(\phi\right)$, where $a$ is the particle diameter. Such fluidity is found to be independent of $\phi$ at low volume fraction~\cite{PoonThomasVriendFranFluid,RHF_PRF2021} but decreases with $\phi$ at sufficiently large volume fraction, vanishing as $\phi$ approaches random close packing, $\phi_{\mathrm{RCP}}$ ~\cite{Phi_rcpFirst}, irrespective of the nature of the flow being homogeneous or inhomogeneous. Moreover, in Ref.~\cite{KimKamrin2020PRL} a new dimensionless number, granular temperature $\Theta = \rho\delta u^2/DP$, is introduced. Using power-law scaling, it is demonstrated that $\mu$ properly scaled by $\Theta$ unifies homogeneous and inhomogeneous rheology with inertial number $I$ being the scaling variable, thus replacing conventional $\mu(I)$ rheology by $\mu(I,\Theta)$ rheology for inhomogeneous flows in dry granular systems. Here, $I$ is the counterpart of $J$ for dry granular system and $D$ is the spatial dimension.         

Similarly, motivated by the concept of $\Theta$ in Ref~\cite{KimKamrin2020PRL}, a recent study~\cite{BPBNessPRL2024} defined a new dimensionless quantity, the suspension temperature $\Theta_s = \eta \delta u/aP$. Using this novel quantity along with the other dimensionless quantities $\phi, J$ and  $\mu$ a set of constitutive laws is identified that unifies the homogeneous and inhomogeneous rheology of dense suspensions of frictionless spherical particles. However, such an approach remains unexplored for frictional and aspherical particles. Interestingly, the constitutive laws for homogeneous rheology differ for frictional and aspherical particles compared to frictionless spherical particles. For frictional particles, as the particle friction coefficient $\mu_p$ increases, the sliding between particle pairs becomes increasingly restricted, imposing constraints on both rotational and translational degrees of freedom. This, in turn, leads to a reduction in $\phi_J^H$~\cite{SinghMariJRheo2018}. With asphericity, the effect on $\phi_J^{H}$ is non-monotonic. Specifically, for rod-shaped particles, $\phi_J^{H}$ decreases when the aspect ratio ($AR = $ length/diameter) exceeds an intermediate value of approximately 1.5. This decrease is due to increased entanglement, or a higher number of contacts per particle, which results in inefficient random packing. However, when the aspect ratio is below this threshold, $\phi_J^{H}$ increases because the reduced entanglement allows for more efficient packing~\cite{MariJRheo,Trulsson_2018,Rod1,Rod2,ChrisPRERod}. Therefore, the validity of $\Theta_s$ for systems of frictional and aspherical particles remains unclear and requires further investigation.

In this work, using particle-based simulations, we unify the homogeneous and inhomogeneous rheology of dense suspensions of frictional spherical particles and frictionless rod-shaped particles. In combination with suspension temperature defined in Ref.~\cite{BPBNessPRL2024} for frictionless spherical particles, along with other dimensionless numbers used to describe the homogeneous rheology, we identify new scaling relations that collapse the data of the homogeneous and inhomogeneous rheology. We find that some of the scaling relations identified here retain the same mathematical form but exhibit different exponent values across systems involving frictionless and frictional spherical particles, as well as frictionless rod-shaped particles. However, other scaling relations apply only to specific systems, highlighting intrinsic differences in their rheology. We further validate these scaling relations by demonstrating their ability to predict various dimensionless numbers in previously unexamined simulation results.   
\begin{figure*}
\includegraphics[width=0.95\textwidth]{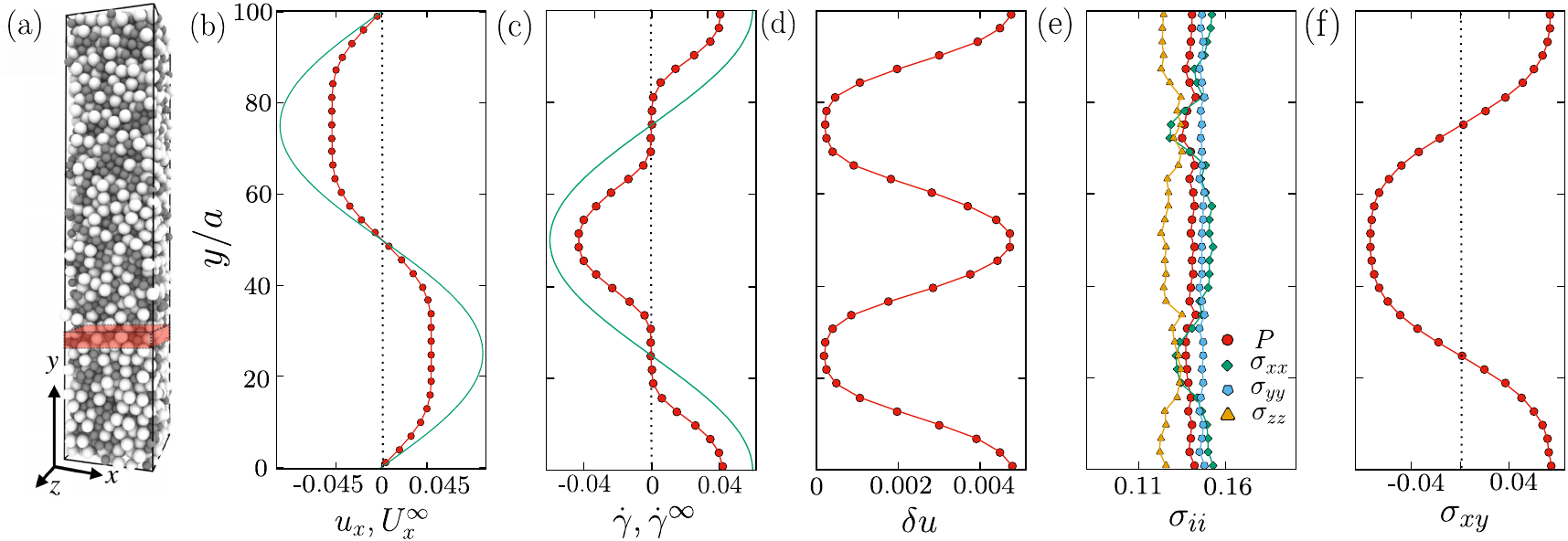}
\caption{Inhomogeneous flow of a dense suspension of frictional spherical particles.
Shown here are
(a) a typical configuration of the system for $\bar{\phi} = 0.60$, with the red region highlighting a coarse-graining box;
and the steady-state profiles in $y$ of
(b) the streaming velocity field $\bm{U}^{\infty}(y)$ in the $\hat{\bm{x}}$ direction (green line) and $x$ component of the coarse-grained velocity field of the particles $\bm{u}$ (red points);
(c) the expected shear rate for a Newtonian fluid $\dot{\gamma}^\infty = \partial U_x^{\infty}/\partial y$ (green line) and the measured shear rate $\dot{\gamma} = \partial u_x/\partial y$ (red points);
(d) the measured velocity fluctuations $\delta u$;
(e) the pressure $P$ and the normal stresses $\sigma_{ii}$;
(f) the shear stress $\sigma_{xy}$ computed from the particle interactions.}
\label{fig0}
\end{figure*}
\section{Simulation details}
We simulate two systems, consisting of non-Brownian frictional spherical particles and frictionless rod-shaped particles using LAMMPS~\cite{LAMMPS,NessSim2023}. For the former, the system is bidisperse, with particle radii $a$ and $1.4a$ mixed in equal numbers to prevent crystallization. For the latter, the rod-shaped particles are created by attaching multiple spheres with appropriate overlap to achieve the desired aspect ratio. Rods constructed in this way behave as rigid bodies, where the forces acting on each sphere within the rod are collectively summed, resulting in a translational force acting on the centre of mass, along with a torque relative to the centre of mass. In both cases, solid particles are suspended in a density $\rho$ matched viscous liquid. The simulations are performed in a periodic box with dimensions $L_x$, $L_y$, and $L_z$ (see Fig.~\ref{fig0}(a)). To vary the solid volume fraction $\phi$, the number of particles is adjusted while keeping the box size constant. To deform the system externally, a space-dependent streaming velocity $\bm{U}^{\infty}({y})$ is introduced. As a result, a solid particle experiences three different types of interactions with its surroundings. First, the drag force and torque on a particle, due to its relative motion with respect to the streaming fluid, are modelled as
\begin{equation}
\bm{f}_i^d = 6\pi\eta a_i\left(\bm U^{\infty}(y) - \bm u_i \right) \text{,}
\label{dragForce}
\end{equation}
\begin{equation}
\bm \tau _i^d = 8\pi\eta a_i^3\left(\bm{\Omega}^{\infty}(y) - \bm{\omega_i} \right) \text{.}
\label{dragTorque}
\end{equation}

Here, $\bm{u}_i$ and $\bm{\omega}_i$ are the linear and angular velocities of the $i^{th}$ particle, and $\bm{\Omega}^{\infty} = \frac{1}{2}\left(\nabla \times \bm{U}^{\infty}\right)$. Second, the presence of viscous fluid resists the relative motion of a pair of particles, modelled here as a hydrodynamic lubrication force~\cite{KimAndKarila,BALL1997444}. The leading-order term of this force and torque between a pair of particles  labelled as $i$ and $j$ with different diameters is given below~\cite{rangarajan_radhakrishnan_2018_1137305,LiRoyerNess2024}

\begin{align}
 \bm f_{i,j}^{h}  \sim \begin{cases} \frac{1}{h^c}\left(\bm{u}_{i,j} \cdot \bm{n}_{i,j}\right)\bm{n}_{i,j} & \text{if $ h_{i,j} < 10^{-3}a'$}; \\
\frac{1}{h_{i,j}} \left(\bm{u}_{i,j} \cdot \bm{n}_{i,j}\right)\bm{n}_{i,j} & \text{if $10^{-3}a' \le h_{i,j} \le 0.5a'$}; \\
0 & \text{if $h_{i,j} > 0.5a'$}.\\
\end{cases}
\label{forceLub}
\end{align}
\begin{align}
 \bm {\tau}_{i,j}^{h}   \sim \begin{cases} \ln \left( \frac{a_i}{h^{c}} \right) \left(\bm{u}_{i,j} \times \bm{n}_{i,j} \right) & \text{if $ h_{i,j} < 10^{-3}a'$}; \\
\ln \left( \frac{a_i}{h_{i,j}} \right) \left(\bm{u}_{i,j} \times \bm{n}_{i,j} \right) & \text{if $10^{-3}a' \le h_{i,j} \le 0.5a'$}; \\
0 & \text{if $h_{i,j} > 0.5a'$}.\\
\end{cases}
\label{forceLub}
\end{align}    
Here $a'$ is the radius of the smaller particle, $h^c = 10^{-3}a'$, $\bm{u}_{i,j} = (\bm{u}_{j} - \bm {u}_{i})$ is the relative velocity. $h_{i,j} = (a_i + a_j) - |r_{i,j}|$ is the shortest distance between surface of two particles, where $a_i$ and $a_j$ are the radius of two different types of particles. $\bm{r}_{i,j}$ is distance between the centres of the two particles pointing from $i^{th}$ to $j^{th}$ particles. $\bm n_{i,j}$ is a unit vector given by $\bm{n}_{i,j} = \bm{r}_{i,j}/|r_{i,j}|$. The reason for making the lubrication force independent of the distance at small $h_{i,j}$ is to allow particles to come into direct contact. Third, the contact force and torque between a pair of particles is modelled in the following way
\begin{align}
 \bm f_{i,j}^{c}  = \begin{cases} k_n h_{i,j}\bm{n}_{i,j}  - k_t \bm{\xi}_{i,j}& \text{if $ h_{i,j} < 0$}; \\
0 & \text{if $ h_{i,j} > 0$}. \\
\end{cases}
\label{forceCont}
\end{align}
\begin{align}
 \bm {\tau}_{i,j}^{c}  = \begin{cases} a_i\left(\bm{n}_{i,j} \times k_t\bm{\xi}_{i,j}\right)& \text{if $ h_{i,j} < 0$}; \\
0 & \text{if $ h_{i,j} > 0$}. \\
\end{cases}
\label{torqueCont}
\end{align}    
$k_n$ and $k_t$ are normal and tangential stiffness chosen as $7\times 10^3$. $\bm{\xi}_{i,j}$ is the accumulated tangential displacement between particles, computed from the time they come into contact until the contact is broken. This displacement accounts for the history dependence of the frictional force~\cite{CundalAndStark}. According to Coulomb’s criterion, the maximum allowable tangential force for frictional particles is given by $k_t\xi_{i,j} < \mu_p k_n h_{i,j}$ . We simulate four different systems of frictional spherical particles with $\mu_p = 0.1,0.2,0.3$ and 0.4 and three different systems of frictionless rod-shaped particles with $AR = 1.5, 2.0$ and 3.0. We obtain homogeneous rheology data for fixed-volume systems over a suitable range of volume fraction by generating simple shear \emph{via}
$\bm{U}^\infty(y)=\dot{\gamma}y\hat{\bm x}$,
with $y$ the direction of the velocity gradient and $\hat{\bm x}$ the unit vector along $x$.
To keep our system in the rate-independent regime, we choose our parameters such that $\rho\dot{\gamma}a^2/\eta\ll1$ and $\dot{\gamma}\sqrt{\rho a^3/k}\ll1$, ~\cite{BoyerGuazelliPouliquenPRL2011}.
To obtain inhomogeneous flow we specify a spatially dependent liquid velocity as $\bm{U}^\infty(y)=\kappa\sin\left(2\pi y/L_y\right)\hat{\bm x}$ (see Fig.~\ref{fig0}(b), and the gradient $\dot{\gamma}^\infty$ in Fig.~\ref{fig0}(c)).  $\kappa$ is a constant with dimension of velocity, chosen to keep $\rho\dot{\gamma}a^2/\eta$ below $0.01$ throughout to be in the overdamped regime. We note that inhomogeneous rheology leads to a spatially varying volume fraction; therefore, in this work, $\phi$ denotes the local volume fraction, while $\bar{\phi}$ represents the mean volume fraction averaged across the entire system. We run simulations with systems containing $\mathcal{O}(10^4)$  particles, and with $\bar{\phi}=0.48$ to $0.62$ for frictional spherical particle system and 0.49 to 0.65 for frictionless rod-shaped particle system.
The stress tensor is computed on a per-particle basis as $\B{\Sigma}_i=\sum_j(\bm{F}_{i,j}^*\otimes\bm{r}_{i,j})$,
counting both contact and hydrodynamic forces.
\begin{figure*}
\includegraphics[width=0.93\textwidth]{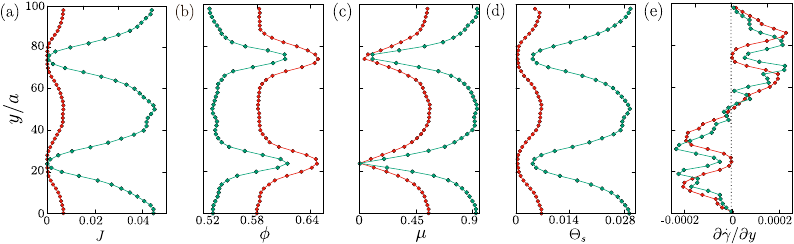}
\caption{The spatial variation of the dimensionless number and the inhomogeneous nature of the flow in the steady state for $\bar{\phi} = 0.60$ (red) and $\bar{\phi} = 0.55$ (green). Shown are (a) the viscous number $J = \eta \dot{\gamma}/P$, (b) the local  volume fraction $\phi$, (c) the macroscopic friction coefficient $\mu = \sigma_{xy}/P$, (d) the suspension temperature $\Theta_s = \eta \delta u/aP$, and (e) the spatial variation of $\partial \dot{\gamma}/\partial y$, where higher values of $\partial \dot{\gamma}/\partial y$ indicates a higher degree of inhomogeneity.}
\label{fig1}
\end{figure*}
To understand the difference between homogeneous and inhomogeneous rheology in our simulation setup, we need to compare the spatially-variant values of $J$, $\mu$, $\phi$ and $\Theta_s$ obtained \emph{via} inhomogeneous flow with the spatially uniform ones obtained \emph{via} homogeneous flow.
In order to do that we compute the variation in $y$ of the stress and velocity fields under inhomogeneous flow,
which we do by binning particle data in blocks of width $a$ and volume $V_b = L_x a L_z$, with the per-block value of a quantity being simply
the mean of the per-particle quantities of the particles with centres lying therein.
We compute the velocity fluctuation
of each particle as $\delta u_i = |u_{i} - \bar{{u}}_{i}|$ where $\bar{{u}}$ is the average velocity of all particles with centres lying in a narrow window $\pm \Delta$ (taking $\Delta=\mathcal{O}(0.1a)$) of $y$, and we then bin $\delta u_i$ per block. To compute $\Theta_s$ we consider only $y$ and $z$ components of $u_i$ to avoid any possible correlated fluctuation originating from the structure in the shear direction (i.e. $\hat{\bm x}$), especially in the case of aspherical particles. The data presented here averaged over approximately 5000 configurations in the steady state.

\section{Results}
In Fig.~\ref{fig0}(b)-(f), steady-state profiles in $y$ of the coarse-grained particle velocity $u_x$ (flow direction), strain rate $\dot{\gamma}=\partial u_x/\partial y$, velocity fluctuations $\delta u$, pressure $P$ ($=(1/3)\text{Tr}(\B{\Sigma})$) and the normal stresses, and the shear stress $\sigma_{xy}$ are shown for frictional spherical particles with $\Bar{\phi} = 0.60$ and $\mu_p = 0.1$, with each plotted point representing a block.
The particle velocity profile and applied streaming velocity follow a similar trend, as expected, but the former is flattened at the regions of largest $\phi$  leading to significant deviations between $\dot{\gamma}$ and $\dot{\gamma}^\infty$ ($=\partial U_x^{\infty}/\partial y$). The pressure becomes spatially uniform in the steady state, and the normal stresses exhibit weak anisotropy at the regions where $\mu > \mu_J$. The shear stress follows similar spatial variation of shear rate.  

In Fig.~\ref{fig1}, the spatial variation of the dimensionless numbers in the steady state is presented for two different $\bar{\phi}$, close to and far from $\phi_J^H$. In Fig.~\ref{fig1}(a), the viscous number $J$ is presented. Since the pressure is uniform in the steady state, $J$ looks similar to $\dot{\gamma}$ shown in Fig.~\ref{fig0}(e). Although we start our simulation from a uniform volume fraction, with straining, particles move towards the region with smaller strain rate due to normal stress $\sigma_{yy}$ imbalance and accumulate there~\cite{migration1,migration2,morris1999curvilinear}. In the steady state $\phi$ attains a maximum at $J =0$ and decreases as $J$ increases. $\mu$ and $\Theta_s$ have a similar variation of $\sigma_{xy}$ and $\delta u$, respectively.
\begin{figure*}
\includegraphics[width=0.97\textwidth]{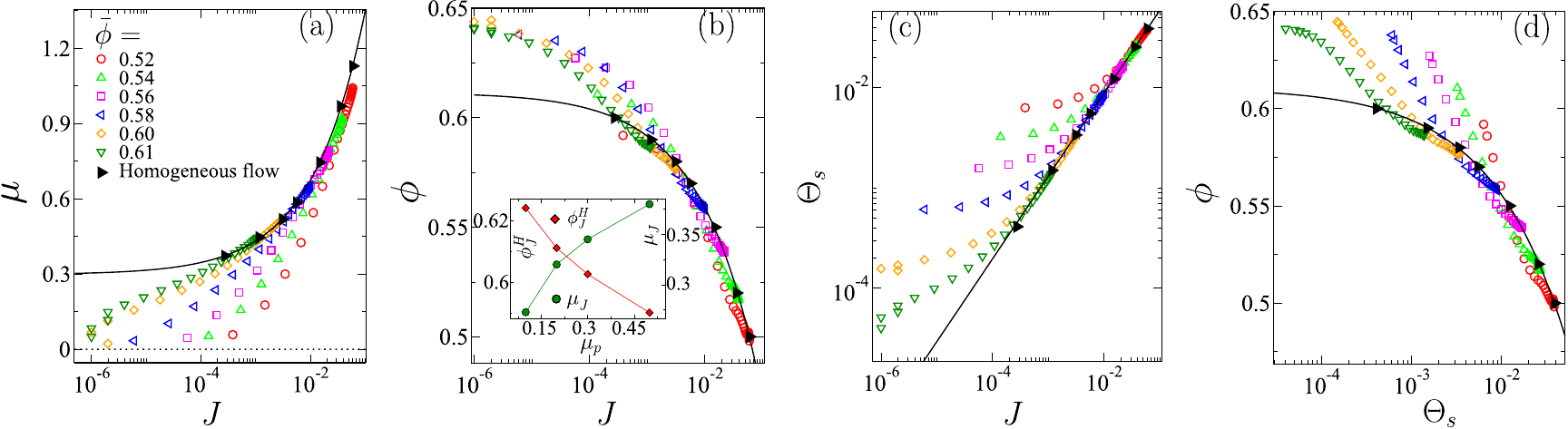}
\caption{Relations between the dimensionless control parameters for a system of frictional spherical particles with $L_x = L_z = 20a$, $L_y = 100a$, and $\mu_p = 0.2$. Shown are the relationships between the dimensionless viscous number $J$ and (a) the macroscopic friction coefficient $\mu$; (b) the local volume fraction $\phi$; and (c) the suspension temperature $\Theta_s$, for a range of mean volume fractions $\bar{\phi}$ in both homogeneous and inhomogeneous flows. In (d), the relationship between $\phi$ and $\Theta_s$ is shown. The inset of (b) shows the dependence of $\phi_J^H$ and $\mu_J$ on $\mu_p$. The black solid lines in (a), (b), (c), and (d) represent the best fits to the homogeneous data based on Eqs.~\ref{eqn1}, \ref{eqn2}, \ref{eqn3}, and \ref{eqn4}, respectively.
}
\label{fig2}
\end{figure*}
\begin{figure*}
\includegraphics[width=0.98\textwidth]{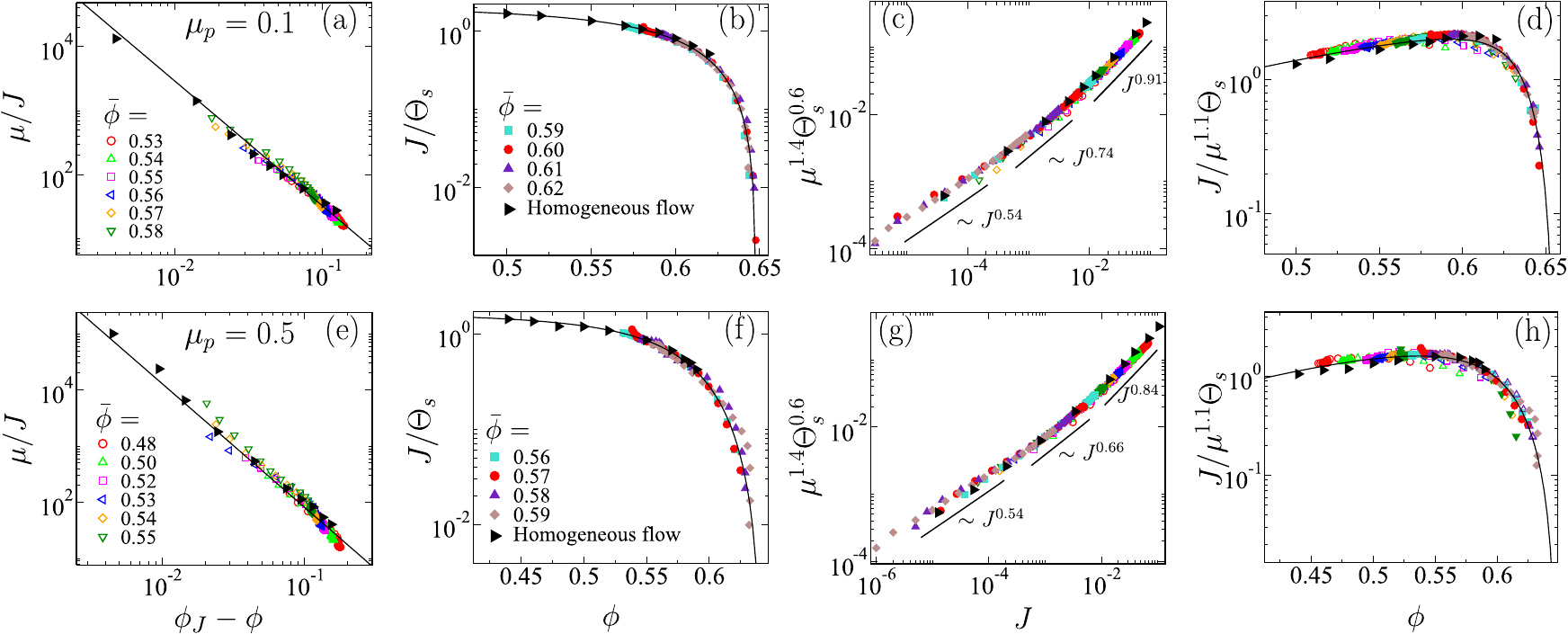}
\caption{Identified scaling relations for frictional spherical particles with two different friction coefficients $\mu_p = 0.1$ (top) and 0.5 (bottom). In (a), (b), (c) and (d), the collapse of homogeneous and inhomogeneous data according to scaling relations given by Eqs.~\ref{scaling1_a}, \ref{scaling1_b}, \ref{scaling2} and \ref{scaling3} for $\mu_p = 0.1$ are presented. The solid black lines represent the scaling function (see text for details). (d)-(g) show the same for $\mu_p = 0.5$.}
\label{fig3}
\end{figure*}
The spatial variation of $\partial \dot{\gamma}/\partial y$, as shown in (e), highlights the inhomogeneous nature of the flow in our setup, as for homogeneous flow, $\partial \dot{\gamma}/\partial y = 0$ throughout. Given that the velocity profile $\bm{U}^{\infty}$ follows a sinusoidal pattern (as shown in Fig.~\ref{fig0}(b)), regions with larger $J$ have smaller $\partial \dot{\gamma}/\partial y$, suggesting that the data from these regions might align with homogeneous (simple shear) flow data. However, regions where both $J$ and $\partial \dot{\gamma}/\partial y$ are small, are less likely to correspond to homogeneous data. These regions exhibit reduced inhomogeneity, as the entire region moves together (creeping motion), as evidenced by the flatness of the velocity profile in Fig.~\ref{fig0}(b). The $\phi$ and $\mu$ here exceeds their homogeneous limit $\phi_J^H$ and $\mu_J$ and the local flow is primarily controlled by $\Theta_s$ as we will see later. Moreover, in Ref.~\cite{BPBNessPRL2024} the inhomogeneity in the flow is quantified for a setup similar to the one studied here by demonstrating a growing length scale associated with the cooperative diffusion of fluidity. This underpins the true inhomogeneous nature of the flow in our setup.          
\subsection{Rheology of frictional spherical particles}
Before we go into the details of identifying scaling relations, we present the dependence of the dimensionless numbers for both homogeneous and inhomogeneous flow in Fig.~\ref{fig2}. In (a), the dependence of the macroscopic friction coefficient $\mu$ on the viscous number $J$ is shown. The black data points represent homogeneous flow and the black solid line is the best fit using a simple power law  form
\begin{equation}
\mu = \mu_J\left(\mu_p\right) + A\left(\mu_p\right)J^{\alpha} \text{.}
\label{eqn1}
\end{equation}
In our study we find $\mu_J$ exhibits strong  dependence on $\mu_p$ as shown in Fig~\ref{fig2}(b) inset, however the exponent $\alpha$ seems to be independent of $\mu_p$ within the studied range. The other data points are for inhomogeneous flow for different mean volume fractions, $\bar{\phi}$. One can clearly see that at large $J$ where the flow is comparatively homogeneous, inhomogeneous data points are superposing on the homogeneous data points. But at comparatively smaller $J$ the effect of inhomogeneity becomes prominent manifesting as the deviation of homogeneous and inhomogeneous data. Moreover, for inhomogeneous flow, $\mu$ goes below $\mu_J$ due to the non-local effect. 

In (b), the dependence of local volume fraction $\phi$ on $J$ is shown. The black points for homogeneous flow, are fitted with the homogeneous constitutive law~\citep{BoyerGuazelliPouliquenPRL2011},
\begin{equation}
\phi = \phi_J^H(\mu_p) - B(\mu_p)J^{\beta} \text{,}
\label{eqn2}
\end{equation}
where $\phi_J^H$ is the $\mu_p$ dependent homogeneous shear jamming volume fraction which is same as $\phi_{{\mathrm{RCP}}}$ for vanishing $\mu_p$ but decreases with increasing $\mu_p$ (see (b) inset and Ref.~\cite{SinghMariJRheo2018}). Similar to the $\mu - J$ plot here also inhomogeneous data fall on homogeneous data for larger $J$ but deviate at smaller $J$. Also, for all inhomogeneous data,  the local volume fraction can be more than $\phi_J^H$ for finite $J$, suggesting that, unlike homogeneous $\mu(J)$ rheology, flow is not solely controlled by $\mu$ and $J$. 

The dependence of $\Theta_s$ on $J$ is shown in (c). For both homogeneous and inhomogeneous flow $\Theta_s$ decreases monotonically with $J$ but at a smaller rate for inhomogeneous flow. For a fixed $J$ inhomogeneous $\Theta_s$ is larger than the homogeneous $\Theta_s$ suggesting a possible significant role of $\Theta_s$ in inhomogeneous flow. For homogeneous flow $\Theta_s$ and $J$ are related by the following power law
\begin{equation}
\Theta_s = G(\mu_p)J^{\gamma} \text{.}
\label{eqn3}
\end{equation} 
In (d), the relation between $\phi$ and $\Theta_s$ is shown. The homogeneous data is fitted with a power law
\begin{equation}
\phi = \phi_J^H\left(\mu_p\right) - D\left(\mu_p \right)\Theta_s^{\delta} \text{.}
\label{eqn4}
\end{equation} 
Similar to other quantities, at smaller $J$ inhomogeneous data deviates from homogeneous data. Interestingly, for $\phi > \phi_J^H(\mu_p)$, $\Theta_s$ remains non-zero indicating the role of $\Theta_s$ in flow in the regions with high $\phi$. For a fixed $\phi$, inhomogeneous flow has higher fluctuations in velocities.\\ 

Our first scaling relation, shown in Fig.~\ref{fig3}(a) and (e), is the divergence of the relative viscosity of the suspension $(\eta_r = \mu/J)$ at the jamming volume fraction, $\phi_J$, given by
\begin{equation}
\mu/J =  \mathcal{F}^S_{1a}\left(\phi\right) \text{,}
\label{scaling1_a}
\end{equation}

with
\begin{equation}
\mathcal{F}_{1a}^S\left(\phi\right) =  \eta_0\left(\mu_p\right)\left(\phi_J\left(\mu_p\right) - \phi\right)^\nu \text{.}
\label{functionF1A}
\end{equation}

Here, $\nu \approx 2$. For homogeneous rheology $\phi_J = \phi_J^H$ monotonically decreases from $\phi_{\text{RCP}}$ with increasing $\mu_p$~\cite{SinghMariJRheo2018}. For inhomogeneous flow, $\phi_J = \phi_J^{IH}$ independent of $\mu_p$ and found to be close to $\phi_{{\mathrm{RCP}}}$. $\eta_0$ is a $\mu_p$ dependent coefficient with different values for homogeneous $\left(\eta_0^H\right)$ and inhomogeneous $\left(\eta_0^{IH}\right)$ rheology. We find $\frac{\eta_0^{IH}}{\eta_0^H} \approx 2$  and $3$ for $\mu_p = 0.1$ and 0.5, respectively. The reason to have such dependence is the following. Since some part of our inhomogeneous simulation box is strained homogeneously (at large $J$) data from this region fall on homogeneous flow data. Thus we have two different power laws for homogeneous and inhomogeneous flow which start from the same point and diverge with the same exponent but at different volume fractions. Therefore, in our scaling relation, we have prefactor that not only depends on the $\mu_p$ but also on the nature of the rheology.

However, for inhomogeneous flow with large $\bar{\phi}$, due to particle migration, the local volume fraction goes above $\phi_J^{H}(\mu_p)$. In this regime, the scaling relation given in Eq.~\ref{scaling1_a} does not apply. The inhomogeneous flow at $\phi > \phi_J^H$ is controlled by $\Theta_s$. In Fig.~\ref{fig2}(b) and (d), the homogeneous and inhomogeneous data do not follow the same trend but for fixed $\phi$, in inhomogeneous flow, both $J$ and $\Theta_s$ seem to have higher values compared to homogeneous values. This indicates that, among the regions which have same $\phi$ flow rate is higher where velocity fluctuations are higher. This correlation leads to another scaling relation presented in Fig.~\ref{fig3}(b) and (f). Here we exploit the power law dependence of $J$ and $\Theta_s$  on $\phi$ given in Eqs.~\ref{eqn2} and~\ref{eqn4} to establish our next scaling relation 
\begin{equation}
J/\Theta_s =  \mathcal{F}^S_{1b}\left(\phi\right)  \text{.}
\label{scaling1_b}
\end{equation}

The data collapse is supported by the following form of scaling function   
\begin{equation}
\mathcal{F}_{1b}^S(\phi) = A_0^S - \frac{A_0^S}{1+A_1^S(\phi_J^{IH} - \phi)^2} \text{,}
\label{functionF1B}
\end{equation}
with $\phi_J^{IH} \sim \phi_{{\mathrm{RCP}}}$. Here, it is important to emphasize that the mathematical form of the scaling function is chosen purely for predictive purposes, without implying any physical significance. While the form of the scaling function may suggest certain physical phenomena, these interpretations may not hold true for the actual system. For example, $\mathcal{F}_{1b}^S(\phi)$ suggests that $J/\Theta_s$ would vanish at $\phi_J^{IH}$ with an exponent of 2; however, this might not be accurate, as we lack data points near $\phi_J^{IH}$ to confirm this behaviour. The primary reason for selecting this specific functional form is that it provides a good fit for data collapse within the studied range. The argument is valid for all the scaling functions used here.     
 
Next we focus on the power law dependence of $\mu$ and $\Theta_s$ on $J$ given by Eqs.~\ref{eqn1} and \ref{eqn3}. In both cases, homogeneous and inhomogeneous data seem scattered but for a fixed $J$ inhomogeneous $\mu$ lie below homogeneous data whereas $\Theta_s$ shows an opposite trend. This suggests that regions with smaller $\mu$ have higher velocity fluctuations which maintains the flow rate. Following Refs.~\cite{KimKamrin2020PRL, BPBNessPRL2024} we attempt to scale $\mu$ by $\Theta_s$ using the power law scaling. From Eqs.~\ref{eqn1} and \ref{eqn3} we expect a power law scaling $\Theta_s \mu \sim J^{\alpha + \gamma}$. However,  unlike dense suspensions of frictionless spherical particles in Ref.~\cite{BPBNessPRL2024}, this power scaling does not result in satisfactory data collapse at small $J$. We find with an adjustment of the weight of $\mu$ and $\Theta_s$ such scaling can provide us with our next scaling relation valid for a wide range of $J$, given below
\begin{equation}
\mu^{1.4}\Theta_s^{0.6} = \mathcal{F}_2^S\left(J\right) \text{.}   
\label{scaling2}
\end{equation}     
Here, $\mathcal{F}_2^S\left(J\right)$ is given by 
\begin{align}
 \mathcal{F}_2^{S}\left(J\right)  = \begin{cases}
C_1^S J^{\alpha_1^S} & \text{if $J > 10^{-2}$}; \\
C_2^S J^{\alpha_2^S} & \text{if $  10^{-2} \ge J > 5\times 10^{-4} $}; \\
C_3^S J^{\alpha_2^S} & \text{if $J \le 5\times 10^{-4}$} \text{.}\\
\end{cases}
\label{functionF2}
\end{align}

The scaling exponents are independent of $\mu_p$ but the form of the scaling function depends on $\mu_p$, with the exponents $\alpha_1^S, \alpha_2^S$ and $\alpha_3^S$ found to be 1, 0.85 and 0.55, and 0.9, 0.75, 0.55 for $\mu_p =$ 0.1  and 0.5, respectively. The data collapse is shown in Fig.~\ref{fig3}(c) and (g). 
\begin{figure*}
\includegraphics[width=0.97\textwidth]{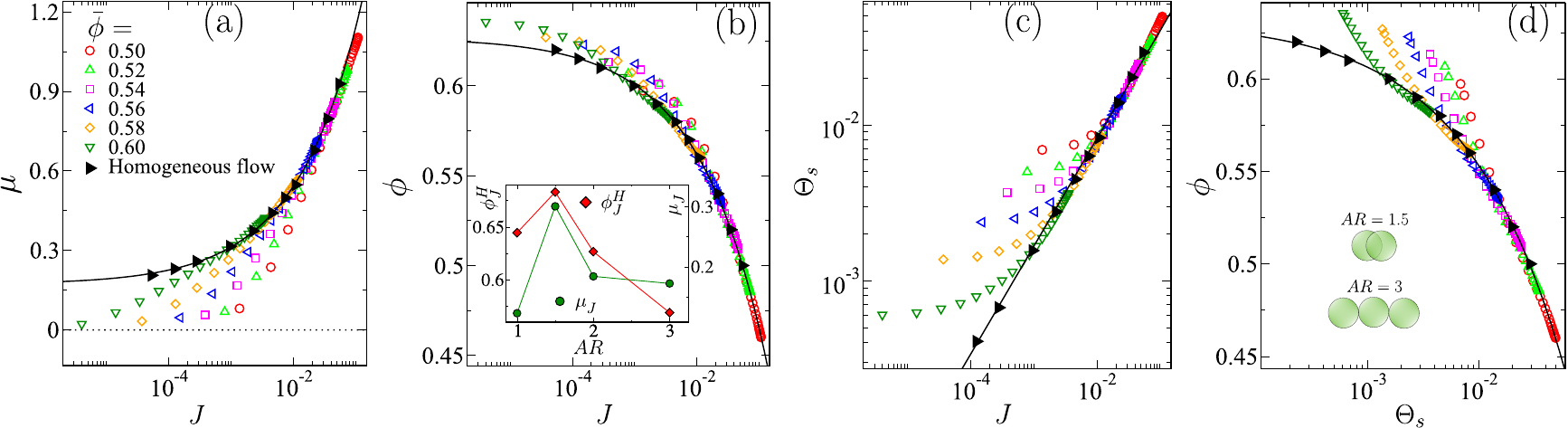}
\caption{The relationships between the dimensionless control parameters for a system of frictionless rod-shaped particles with $L_x = L_z = 25a$, $L_y = 80a$, and aspect ratio $AR = 2.0$. Shown are the relationships between the dimensionless viscous number $J$ and (a) the macroscopic friction coefficient $\mu$; (b) the local volume fraction $\phi$; and (c) the suspension temperature $\Theta_s$, for a range of mean volume fractions $\bar{\phi}$ in both homogeneous and inhomogeneous flows. In (d), the relationship between $\phi$ and $\Theta_s$ is shown. The inset of (b) shows the dependence of $\phi_J^H$ and $\mu_J$ on $AR$. The black solid lines in (a), (b), (c), and (d) represent the best fits to the homogeneous data based on Eqs.~\ref{eqn1}, \ref{eqn2}, \ref{eqn3}, and \ref{eqn4}, respectively. The inset of (d) shows schematics of rods with two aspect ratios, $AR = 1.5$ and 3.}
\label{fig4}
\end{figure*}
\begin{figure*}
\includegraphics[width=0.97\textwidth]{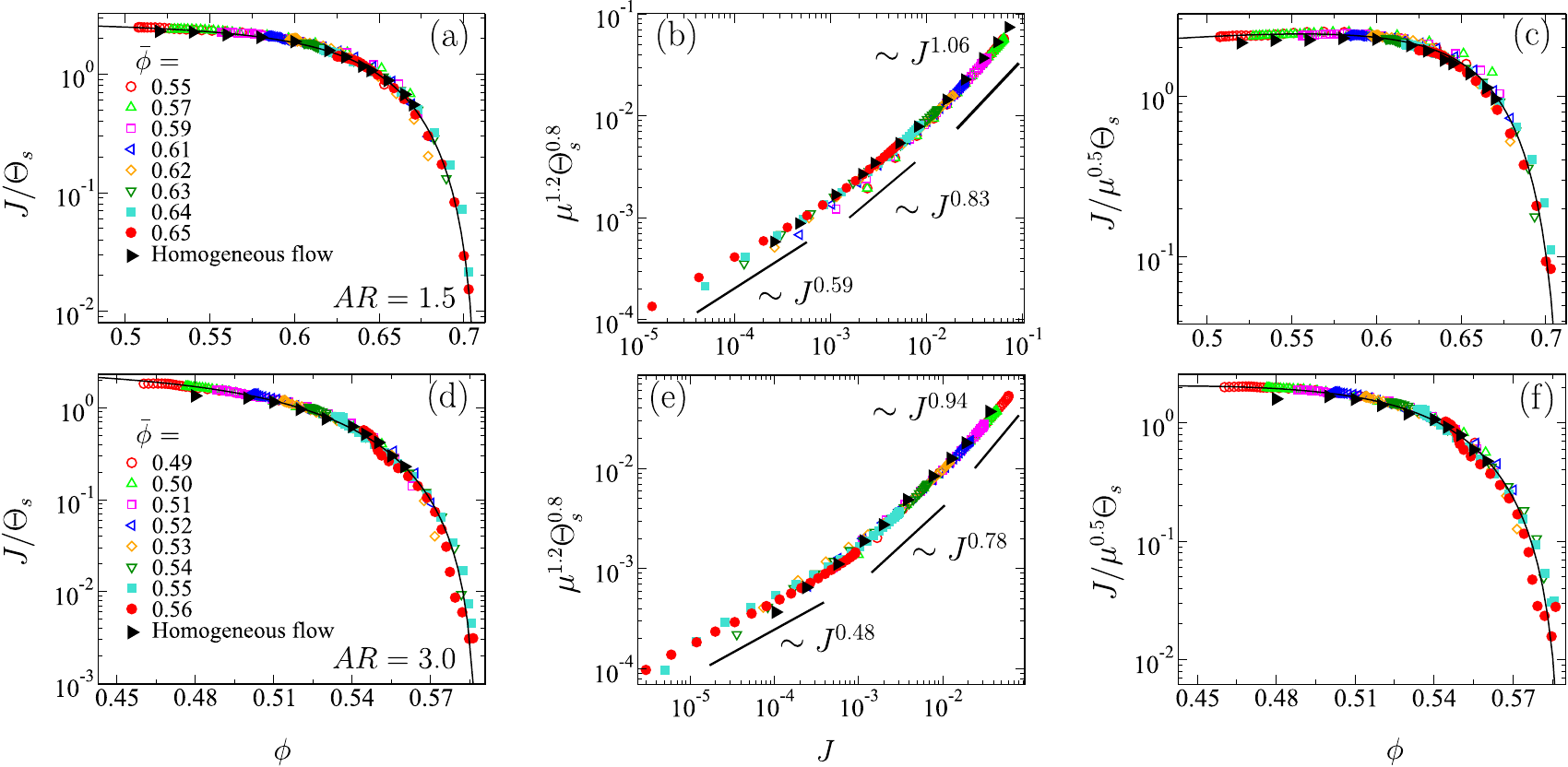}
\caption{Identified scaling relations for frictionless rod-shaped particles with two different aspect ratios, $AR = 1.5$ and 3. In (a), (b) and (c) the collapse of homogeneous and inhomogeneous data according to scaling relations given by Eqs.~\ref{scaling5}, \ref{scaling6} and \ref{scaling7} are shown. The solid black lines represent the scaling function (see text for details). (d)-(f) show the same for $AR = 3$.}
\label{fig5}
\end{figure*}

\begin{figure*}
\includegraphics[width=0.97\textwidth]{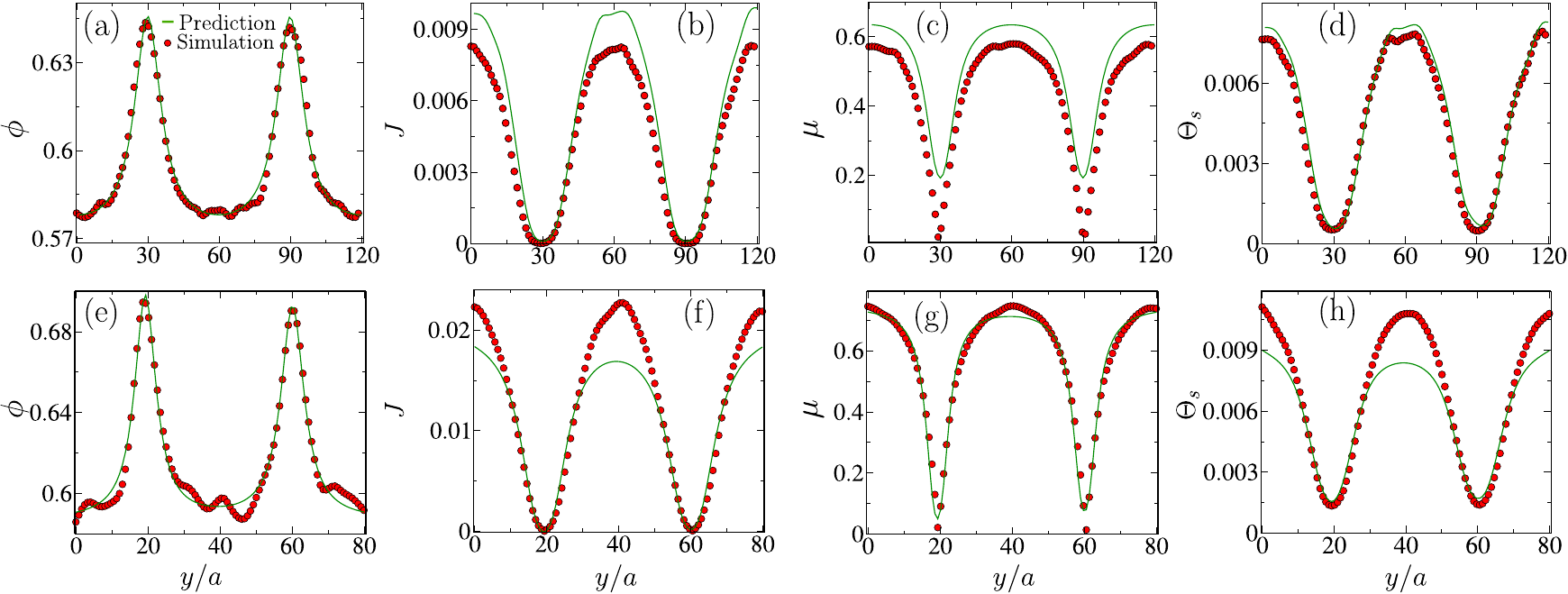}
\caption{Predictions of spatial variation of the dimensionless quantities $\phi$, $J$, $\mu$ and $\Theta_s$ using the identified scaling relations and momentum balance equation against simulation data not used for the data collapse, for the system of frictional spherical particles (top) and frictionless rod-shaped particles (bottom) with $\bm{U}^\infty(y)=\kappa\sin(2\pi y/L_y)\hat{\bm x}$.
Shown are (a) the volume fraction $\phi$;
(b) the viscous number $J$;
(c) the macroscopic friction coefficient $\mu$;
and (d) the suspension temperature $\Theta_s$,
with predictions given by solid green lines and simulation data in red points, for $\Bar{\phi} = 0.595$ and $\mu_p = 0.1$. The same quantities are shown in (e), (f), (g) and (h) for the system of frictionless rods for $\Bar{\phi} = 0.615$ and $AR = 1.5$.}
\label{fig6}
\end{figure*}
Our next scaling relation is based on the granular fluidity defined as $J/\mu \Theta$ (see Ref.~\cite{ZhangKamrinPRL2017}) which uniquely depends on $\phi$ for both homogeneous and inhomogeneous flows. The theoretical justification of such quantity is given by kinetic theory \cite{KineticTheory1,KineticTheory2}. Similar to this an effective suspension fluidity is introduced in Ref.~\cite{BPBNessPRL2024} for suspension of spherical frictionless particles. Here we extend this to the suspension of frictional spherical particles which gives us 
\begin{equation}
J/\mu^{1.1}\Theta_s = \mathcal{F}_3^{S}\left(\phi\right) \text{.}   
\label{scaling3}
\end{equation}
Here
\begin{equation}
\mathcal{F}_3^S(\phi) = D_0^S\phi^{3} - \frac{D_0^S \phi^{3}}{1+D_1^S(\phi_J^{IH} - \phi)^2} \text{,}
\label{functionF3}
\end{equation}
shown in Fig.~\ref{fig3}(d) and (h).\\
      
Thus, we effectively have three scaling relations. The second and third scaling relations, given by Eqs.~\ref{scaling2} and ~\ref{scaling3}, are valid across a wide range of volume fractions, both above and below $\phi_J^{H}$. In contrast, the first scaling relation is divided into two parts. The first part, Eq.~\ref{scaling1_a}, applies for $\phi < \phi_{J}^{H}$, while the second part, Eq.~\ref{scaling1_b}, is valid above $\phi_J^{H}$. Later, we will demonstrate how these scaling relations, combined with the momentum balance equation, allow us to predict all the relevant dimensionless numbers based solely on the applied fluid flow.

\subsection{Rheology of frictionless rod-shaped particles}
For the system of frictional spherical particles we find decoupling of homogeneous and inhomogeneous shear jamming volume fraction (i.e. $\phi_J^H < \phi_{\mathrm{RCP}} \approx \phi_J^{IH}$) due to the frictional constraints. Similar decoupling is also expected for rod-shaped particle due to the constraints imposed by asphericity quantified by the aspect ratio. The relations between different dimensionless numbers for this system, shown in Fig.~\ref{fig4}, are similar to those for frictional spherical particles. The homogeneous data follow the same functional form given by Eqs.~\ref{eqn1}, \ref{eqn2}, \ref{eqn3} and \ref{eqn4}, and are therefore not discussed here. Both $\phi_J^H$ and $\mu_J$ exhibit non-monotonic dependence on $AR$, with an maximum value of $AR \approx 1.5$, shown in inset of Fig.~\ref{fig4}(b).

We further identify the scaling relations to unify the homogeneous and inhomogeneous flow of rod-shaped, frictionless particles. We find that the scaling relation $\mu/J \sim \mathcal{F}_{1a}^{S}(\phi)$ which works for frictionless spherical particles across the entire range of volume fractions and for frictional spherical particles within a limited range, particularly below $\phi_J^H$, does not hold for frictionless rod-shaped particles. However, the scaling relation in Eq.~\ref{scaling1_b} holds over a wide range of volume fractions, including both above and below the aspect ratio-dependent $\phi_J^H$. This is our first scaling relation for the system of frictionless of rod-shaped particles, which can be expressed as
\begin{equation}
J/\Theta_s = \mathcal{F}_1^R\left(\phi\right) \text{.}
\label{scaling5}
\end{equation}
Here 
\begin{equation}
\mathcal{F}_1^R\left(\phi\right) = A_0^R - \frac{A_0^R}{1+A_1^R(\phi_J^{IH} - \phi)^{2.4}}
\label{functionR1}
\end{equation}
is the best fitted form of the master curve which vanishes at $\phi_J^{IH}$ as shown in Fig.~\ref{fig5}(a) and (d), for $AR$ = 1.5 and 3.0. Unlike, the frictional spherical system, for rod-shaped particles $\phi_J^{IH}$ can be different from $\phi_{{\mathrm{RCP}}}$.

Similar to the scaling relation given by Eq.~\ref{scaling2} for the system of frictional spherical particles,  we find such power law scaling also works for rod-shaped particles but with different exponents. 
\begin{equation}
\mu^{1.2}\Theta_s^{0.8} = \mathcal{F}_2^{R}\left(J\right) \text{,}   
\label{scaling6}
\end{equation}
where the best form of the scaling function $\mathcal{F}_2^{R}\left(J\right)$ is the following. 
\begin{align}
 \mathcal{F}_2^{R}\left(J\right)  = \begin{cases}
C_1^R J^{\alpha_1^R} & \text{if $ J > 2 \times 10^{-2}$}; \\
C_2^R J^{\alpha_2^R} & \text{if $ 2 \times 10^{-2} \ge J > 8\times 10^{-4}$}; \\
C_3^R J^{\alpha_3^R} & \text{if $J \le 8\times10^{-4}$} \text{.}\\
\end{cases}
\label{functionR2}
\end{align}
As with frictional spherical particles, the scaling exponents are independent of $AR$ but the exponents $\alpha_1^R, \alpha_2^R$, and $\alpha_3^R$ in the master curve are found to be 1.06, 0.83, and 0.59 for $AR = 1.5$, and 0.94, 0.78, and 0.48 for $AR = 3.0$, respectively, see Fig.~\ref{fig5}(b) and (e).

Similar to frictional spherical particles our third scaling relation for rod-shaped particles describes the dependence on effective suspension fluidity on the volume fraction uniquely defined for both homogeneous and inhomogeneous rheology. We find in this system the effective suspension fluidity can be defined as $J/\mu^{0.5}\Theta_s$ and exhibits the following relation as shown in Fig.~\ref{fig5}(c) and (f).
\begin{equation}
J/\mu^{0.5}\Theta_s = \mathcal{F}_3^{R}\left(\phi\right) \text{,}   
\label{scaling7} 
\end{equation}
where 
\begin{equation}
\mathcal{F}_3^{R}\left(\phi\right) = D_0^{R}\phi - \frac{D_0^{R}\phi}{1 + D_1^{R}(\phi_J^{IH} - \phi)^2} \text{,}
\end{equation}

vanishing at $AR$ dependent volume fraction $\phi_J^{IH}$.             \\

\section{Prediction}
Our system is characterised by four dimensionless numbers and three effective scaling relations. By examining the spatial variation of just one of these dimensionless numbers, we can comprehensively capture and describe the system’s rheological behaviour. In our simulations, however, the only known input is the externally applied streaming velocity profile, represented by $\bm{U}^\infty (y)$.  Thus, to utilize the scaling relations we must be able to compute one of the dimensionless numbers from the information of $\bm U^{\infty}$. To do so, considering the inertia-free momentum balance $\nabla\cdot\bm{\Sigma} = -\bm{f}$ per unit volume, for the $l^{th}$ segment of the simulation cell we express the following equation
\begin{equation}
N_l6\pi\eta a_l\left[{U}_{x,l}^\infty - {u}_{x,l} \right] = -\left( \frac{\partial\sigma_{xy,l}}{\partial y} \right) V_b\text{.}
\label{momBalance}
\end{equation}

$N_l$, $U_{x,l}^\infty$, $u_{x,l}$, and $\sigma_{xy,l}$ represent the particle number in the block, the liquid streaming velocity at the block centre, the particle velocity, and the stress averaged over the block, which has volume $V_b$. $a_l$ represents a volume averaged particle radius at $l$ with magnitude $\approx 1.2$ for spheres and $1$ for rod-shaped particles. On the left hand side of Eq.~\ref{momBalance}, the first term denotes the net applied external force, while the second term signifies the net viscous force due to fluid drag. The difference between these forces is compensated by the net stress gradient within the block. Using our dimensionless number definitions, Eq.~\ref{momBalance} can be reformulated for the streaming velocity at $y$ as

\begin{equation}
{U}_{x}^{\prime\infty}(y) = \left[\int_0^y\frac{1}{a}J^{\text{*}}(y')dy' - \frac{2a}{9\phi(y)}\left( \frac{\partial\mu^{\text{*}}(y)}{\partial y} \right)\right]\text{,}
\label{momBalanceDimLess}
\end{equation}
where ${U}_{x}^{\prime\infty}(y)={U}_{x}^{\infty}(y)\eta /aP$ and the asterisks indicate multiplication by $\mathrm{sgn}(\dot{\gamma}^\infty(y))$. Note that $P$ is uniform at steady state and $\phi(y) = (4/3)\pi a^3 N(y)/V_b$. Thus, Eq.~\ref{momBalanceDimLess} links the externally applied liquid flow field to the profiles of $J$, $\mu$, and $\phi$.

Given ${U}^\infty_x$, we solve Eq.~\ref{momBalanceDimLess} and the scaling relations (Eqs.~\ref{scaling1_a}, \ref{scaling1_b}, \ref{scaling2} and  \ref{scaling3} for frictional spherical particles, and Eqs.~\ref{scaling5}, \ref{scaling6} and \ref{scaling7} for frictionless rod-shaped particles) numerically in the following method. Initially, we assume a $\phi\left(y\right)$ profile, hypothesizing accumulation of particles at points where the spatial derivative of $\bm{U}^\infty$ is zero, starting with a simple Lorentzian form $\phi(y) = \sum_{k=1}^{n_p} a_{k}/[(y - y^0_k)^2 + b^2_{k}] + \phi_0$\text{,} with mass conserved through $\bar{\phi} = \frac{1}{L_y}\int_0^{L_y} \phi\left(y\right)dy$. Here, $n_p$ is the number of points where the first derivative of $\bm{U}^\infty$ is zero, $y^0_k$ is the coordinate of such a point, and $b_k$ is the width of the Lorentzian function centered at $y^0_k$. Next, we compute $J$, $\mu$, and $\Theta_s$ directly using the scaling relations, before attempting to balance Eq.~\ref{momBalanceDimLess}. The imbalance in Eq.~\ref{momBalanceDimLess} indicates the accuracy of our initial guess. We refine $\phi(y)$ by adjusting $\phi_0$, $a_k$, and $b_k$ until Eq.~\ref{momBalanceDimLess} is satisfied within an acceptable tolerance.   

The results, shown in Fig.~\ref{fig6}, compare predicted outcomes against previously unseen simulation data (i.e., data not used for obtaining the scaling exponents) with $\bar{\phi}=0.595$ for frictional spherical  particles, $0.615$ for frictionless rod-shaped particles, and $\bm{U}^\infty(y)=\kappa\sin(2\pi y/L_y)\hat{\bm x}$, demonstrating the success scaling relations in predicting $y$ profiles of $\phi$, $J$, $\mu$, and $\Theta_s$. Despite the highly non-linear nature of the scaling relations and many orders spread of $J$ and $\Theta_s$, the predictions are reasonably accurate.

\section{Conclusion}
Through particle-based simulations, we establish a universal description of the flow behaviour of dense suspensions, of frictional spherical and frictionless rod-shaped particles. In addition to the standard control parameters solid volume fraction $\phi$, viscous number $J$, and macroscopic friction coefficient $\mu$, we introduce a novel parameter, suspension temperature $\Theta_s$, representing velocity fluctuations, inspired by concepts from dry granular materials. Our findings reveal scaling relations among these parameters that successfully collapse data for both homogeneous and inhomogeneous flows. Using the momentum balance, we demonstrate that the characteristics of general homogeneous and inhomogeneous flow can be predicted based on the applied external force. It is important to note that in this work we primarily focused on finding the existing scaling relations from the simulation data rather than a thorough investigation of the physical origin of them. The exponents reported here are found by using an ad-hoc method to obtain the best data collapse. The physical origin of these exponents and the validity of the identified scaling relations in flows with more complex geometries remain unexplored here, but represent natural and important avenues for future work.

\section{Acknowledgment}
B.P.B. acknowledges support from the Leverhulme Trust under Research Project Grant No. RPG-2022-095; C.N. acknowledges support from the Royal Academy of Engineering under the Research Fellowship scheme. We are grateful to Anoop Mutneja, Eric Breard and Ken Kamrin for discussions. 
  
\bibliography{ALL1}
\end{document}